\documentclass[pre,twocolumn, showpacs,superscriptaddress,amsmath,amssymb]{revtex4}
\usepackage{graphicx}
\usepackage{amssymb}
\usepackage{amsmath}
\begin{document}
\title{Alternative method for wave propagation analysis within
bounded linear media: conceptual and practical implications}

\author{Alberto Lencina}
\email[Corresponding author. Electronic address: ]
{agl@fisica.ufpb.br} \affiliation{Departamento de F\'isica, Centro
de Ci\^encias Exatas e da Natureza, Universidade Federal da
Para\'iba, Caixa Postal 5008 CEP 58051-970, Jo\~ao Pessoa,
Brazil.}
\author{Beatriz Ruiz}

\affiliation{Centro de Investigaciones \'Opticas, cc 124 (1900),
La Plata, Argentina.}

\author{Pablo Vaveliuk}

\affiliation{Departamento de F\'isica, Universidade Estadual de
Feira de Santana, Campus Universit\'ario BR 166 KM 03, 44031-460 Feira de Santana, Bahia, Brazil.}

\begin{abstract}
This paper uses an alternative approach to study the monochromatic
plane wave propagation within dielectric and conductor linear
media of plane-parallel-faces. This approach introduces the
time-averaged Poynting vector modulus as field variable. The
conceptual implications of this formalism are that the
nonequivalence between the time-averaged Poynting vector and the
squared-field amplitude modulus is naturally manifested as a
consequence of interface effects. Also, two practical implications
are considered: first, the exact transmittance is compared with
that given by the Beer's Law, employed commonly in experiments.
The departure among them can be significative for certain material
parameter values. Second, when the exact reflectance is studied
for negative permittivity slabs, it is show that the high
reflectance can be diminished if a small amount of absorption is
present.
\end{abstract}
\pacs{41.20.Jb, 42.25.Bs, 41.85.Ew}
\date{\today}
\maketitle

\section{Introduction}
The counter-propagating wave approach is commonly used to study
the electromagnetic response of spatially nondispersive,
homogeneous and isotropic plane-parallel-faces linear media
(Fabry-Perot framework). It consider forward and backward
monochromatic plane waves to derive the optical properties as the
transmittance and reflectance \cite{straton}. This method uses the
conventional variables of electromagnetic fields: moduli and
phases of these waves which should be found from the Helmholtz
equation with the corresponding boundary conditions. It is
well-known that for a single harmonic plane wave propagating
through an unbounded linear medium, the time-averaged Poynting
vector modulus is equivalent to the squared-field amplitude
modulus \cite{jackson}. This statement is generally accepted even
if the wave propagation takes place in bounded linear media. The
counter-propagating approach does not point out the possible
nonequivalence between the above magnitudes since utilizes the
amplitudes and phases as field variables. However, the phase of
the wave can be replaced by the time-averaged Poynting vector
modulus as field variable. Then the time-averaged Poynting vector
and the squared-field amplitude modulus can be monitored
simultaneously within the medium so that the conditions that leads
to the nonequivalence among them will manifest naturally in this
equivalent frame.

The so-called S-formalism use the time-averaged Pointing vectos as
a field variable. It was recently introduced \cite{pre} and
applied to study the optical response of nonlinear slabs. The
nonequivalence between the time-averaged Poynting vector and the
squared-field amplitude modulus was the key to define a nonlinear
medium whose nonlinearity is proportional to the time-averaged
Poynting vector modulus. Its transmittance was calculated and
found to differ with that obtained for the Kerr medium, whose
nonlinearity is proportional to the squared-field amplitude
modulus. However, the method was not yet used to analyze the
linear case. Therefore, the aim of this paper is to apply the
S-Formalism to study the optical properties of homogeneous,
isotropic and spatially non-dispersive dielectric as well as Ohmic
conductor media, inside the Fabry-Perot frame complementing to the
well-known results of this problem within the counter-propagating
wave approach. The S-formalism shows, in direct form, how this
non-equivalence is related to the superposition dynamic of
opposite traveling waves which is only possible in bounded media.
Also, the exact transmittance is compared with that given by the
approximated Beer's Law, commonly used in experiments, and
appreciable differences were found for certain material parameters
values. Moreover, the reflectance of positive and negative
permittivity media are considered and interesting results appears
for small absorption when negative permittivity accounts.

It must be stressed that the S-formalism does not substitute the
counter-propagating wave approach because the idea of forward and
backward waves within the medium has a deep physical content.
However, the alternative method adds conceptual and practical
implications that remains hidden when the conventional scheme is
utilized. We believe that both methods should be complementary,
helping for a complete physical understanding on wave propagation
in bounded linear media.

Below, this paper reviews the principal features of the
S-Formalism introduced in Ref. \onlinecite{pre} as function of the
new set of field variables. Then, it is applied to derive the
spatial evolution of both variables for absorbent and nonabsorbent
linear media showing the cases in which the equivalence does not
hold true, and their causes and consequences. Finally, practical
implications of this features are explored: the comparison between
the exact transmittance and the Beer's law, and the analysis of
positive and negative permittivity media.

\section{S-Formalism approach}

\begin{figure}[t]
\includegraphics[width=7.5cm,height=5.cm]{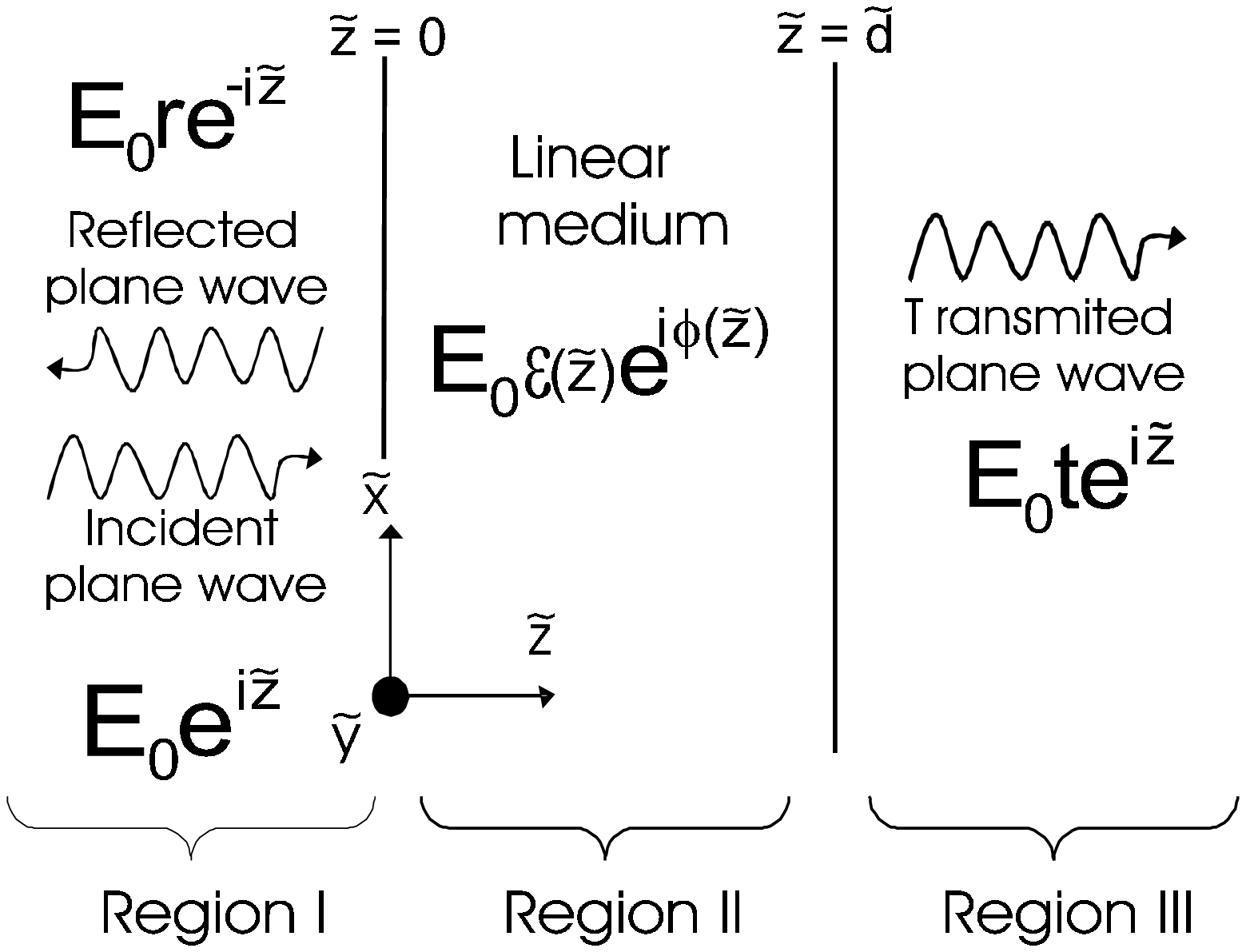}
\caption{A harmonic plane wave of amplitude $E_0$, wave vector
$k_0$ and frequency $\omega$ strikes a linear, non-magnetic,
homogeneous, isotropic, and spatially non-dispersive
plane-parallel-faces, to be reflected and transmitted. The field
in Region II can be represented by a general complex field of real
amplitude $\mathcal{E}$ and phase $\phi$. The Regions I and III
constitute, for simplicity, the same linear dielectric medium
(e.g., air). $r$ and $t$ are the reflection and transmission
complex coefficients, respectively; $\tilde{z}=k_0 z$ is the
dimensionless propagation coordinate.} \label{fig=problema}
\end{figure}

When a linear, non-magnetic, homogeneous, isotropic, and spatially
non-dispersive plane-parallel-faces medium of dimensionless
thickness $\tilde{d} \: (=k_0 d)$ is
excited perpendicularly by a plane wave (see Fig. \ref{fig=problema}), the
S-Formalism \cite{pre} states that the transmittance and
reflectance are respectively
\begin{subequations}
\begin{eqnarray}
R=|r|^2=1-S(0),\\
T=|t|^2=S(\tilde{d}),
\end{eqnarray}
\end{subequations}
 where $S$ is the field variable, representing the
  dimensionless time-averaged Poynting vector modulus, given by
 \begin{eqnarray}
S(\tilde{z})=\frac{\left\langle
\mathbf{S}\right\rangle\cdot\mathbf{\hat{e}}_z}{ I_0} =
\mathcal{E}^2 \frac{d\phi}{d\tilde{z}} \label{sye}
\end{eqnarray}
 where $I_0=\epsilon_0\omega/(2k_0)\:E_0^2$ is the incident
intensity with $\epsilon_0$, the vacuum permittivity,
$\mathbf{\hat{e}}_z$ the unit vector in the $z$ direction, and
$\mathcal{E}$ and $\phi$ are the electric field amplitude modulus
and phase, respectively. $S(\tilde{z})$ and
$\mathcal{E}(\tilde{z})$ evolve according to the coupled system
\cite{pre}
\begin{subequations}
\begin{eqnarray}
&&\mathcal{E}^3\frac{d^2\mathcal{E}}{d \tilde{z}^2}+\epsilon_r
\:\mathcal{E}^4
-S^{2}=0,\label{fs1}\\[3pt]
&&\frac{d S}{d \tilde{z}}+\sigma_r\:\mathcal{E}^{2}
\label{eq=poyntinhtheo}=0,
\end{eqnarray}
\label{eq=formalismoS}
\end{subequations}
with boundary conditions
\begin{subequations}
\begin{eqnarray}
&&\left[\left(\mathcal{E}+\frac{S}{\mathcal{E}} \right)^{2}+\left(
\frac{d\mathcal{E}}{d \tilde{z}}
\right)^{2}\right]_{\tilde{z}=0}=4,\label{eq=condecont1}\\[3pt]
&&\left[S-\mathcal{E}^{2}\right]_{\tilde{z}=\tilde{d}}=0,\\[3pt]
&&\left[\frac{d\mathcal{E}}{d
\tilde{z}}\right]_{\tilde{z}=\tilde{d}}=0.
\end{eqnarray}\label{eq=condecont}
\end{subequations}
being $\epsilon_r=\epsilon/\epsilon_0$, the relative permittivity,
and $\sigma_r=\sigma/\left (\epsilon_0 \omega \right)$, the
relative conductivity. From Eq. (\ref{sye}) it is clear that $S$
and $\mathcal{E}$ are equivalents only when $\phi$ is a linear
function on $z$. This point is
 not, in general, emphasized in the literature.
Thereby, it is mandatory to ask: is the equivalence of both
magnitudes generalized to cases where it could be not longer
true?. An example: in Ref. \cite{gb}, the energy flux of a
TEM$_{00}$ Gaussian beam propagating in free-space is calculated
from the squared-field amplitude modulus instead of the
time-averaged Poynting vector which could lead to non-physical
results. For details see Appendix \ref{appendix}.

Equation (\ref{eq=poyntinhtheo}) represents the time-averaged
Poynting Theorem being $\sigma_r$ the responsible for the energy
loss in the medium. Here we used it \emph{explicitly} to analyze
and solve wave propagation problems in bounded media. On the other
hand, note that, Eqs. (\ref{eq=condecont}) are independent of the
medium properties. Also, the second interface produces different
values of $\mathcal{E}^2$ and $S$ at the first interface through
the spatial dependence of $\mathcal{E}$, as can be seen in Eq.
(\ref{eq=condecont1}). However at the second interface
$\mathcal{E}^2$ and $S$ are always equals with the advertisement
that this should not lead to false view that they are equivalents.

Finally note that, the energy conservation is guaranteed thorough
the expression:
\begin{eqnarray}
|r|^2+|t|^2=1-\left[S(0)-S(\tilde{d})\right].
\end{eqnarray}
showing that the reflectance and transmittance values are limited
by the boundary values of $S$ and enhancing their importance in
wave propagation problems in bounded media.

\section{Wave propagation analysis}
\subsection{Linear dielectric}

The linear dielectric is the simplest nonabsorbent medium. It is
clear from Eq. (\ref{eq=poyntinhtheo}) that the dimensionless
intensity $S$ is a constant fixed by the boundary conditions when
$\sigma_r=0$. In this case, the analytical solutions for $S$ and
$\mathcal{E}^2$ are
\begin{subequations}
\begin{eqnarray}
S&=&\frac{1}
{1+F \sin^2\left(\delta/2\right)},\label{eq=bcPoy}\\[5pt]
\mathcal{E}^2(\tilde{z}_r)&=&\frac{1-\left(4F/(\epsilon_r-1)\right)
\sin^2\left[(1-\tilde{z}_r)\delta/2\right]}{1+F
\sin^2\left(\delta/2\right)}, \label{eq=epsPoy}
\end{eqnarray}
\label{eq=soldielectrico}
\end{subequations}
\noindent where $F=(\epsilon_r-1)^2/(4\epsilon_r)$ is known as
\emph{finesse}, $\delta=2 \tilde{d} \sqrt{\epsilon_r}$ and
$\tilde{z}_r=\tilde{z}/\tilde{d}$.
\begin{figure}[t]
\includegraphics[width=7.5cm,height=5.cm]{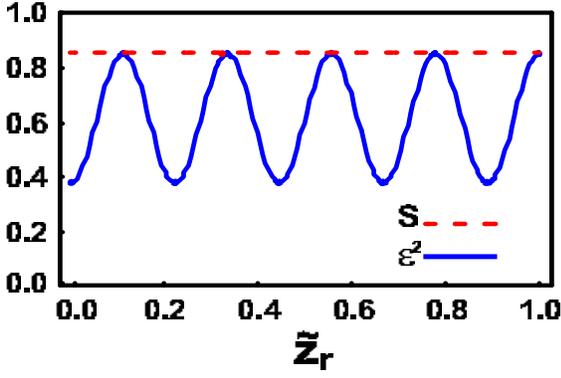}
\caption{(Color online) Spatial evolution of the dimensionless
intensity $S$ and squared-field amplitude modulus $\mathcal{E}^2$
for a linear dielectric. Clearly, $S$ is a constant and
$\mathcal{E}^2$ is an oscillatory function. The parameter values
are $\sqrt{\epsilon_r}=1.5$, $k_0d=3\pi$.} \label{fig=dielectrico}
\end{figure}

Equation (\ref{eq=bcPoy}) is the transmittance result for the
linear dielectric with Fabry-Perot geometry at normal incidence,
i. e. the well-known Airy-formula \cite{wolf}. The solutions
(\ref{eq=soldielectrico}) explicitly show that the intensity is
non-equivalent to the squared-field amplitude modulus. Figure
\ref{fig=dielectrico} points out the nonequivalence. The intensity
is a constant within the medium while the squared-field amplitude
is an oscillatory function. Note that $\mathcal{E}$ and $S$ are
only equals at $\tilde{z}_r=1-2m \pi  /\delta$ with
$m\:\varepsilon\: \mathbb{N}$. Observe that when $\tilde{d}=m'\pi
/\sqrt{\epsilon_r}$ ($m'\:\varepsilon\: \mathbb{N}$),
$\delta=2m'\pi$, then $\mathcal{E}(0)=S=1$ attaining their maximum
values ($T=1$ and $R=0$). In this case the medium behaves as a
\textit{delay sheet} transmitting all the incident intensity. When
$\delta=(2m'+1)\pi$ the field amplitude modulus and the
temporal-averaged Poynting vector modulus attain their minimum
values given by $\mathcal{E}(0)=[1-4F/(\epsilon_r-1)]/(1+F)$ and
$S=1/(1+F)$. Note that, when $\epsilon_r$ increases, F also
increases and the minimum of $S$ diminishes.

On the other hand, when are both magnitudes really equivalents?
When the squared-field amplitude modulus is a constant. In this
case Eqs. (\ref{eq=formalismoS}) relate $S$ and $\mathcal{E}$ by
\begin{eqnarray}
S=\sqrt{\epsilon_r}\:\mathcal{E}^2.\label{eq=equivdiel}
\end{eqnarray}
This case corresponds to the \emph{\textbf{single}} plane wave
propagating in an dielectric medium \cite{jackson}.

\subsection{Linear absorber}

The linear absorber or Ohmic conductor medium is characterized by
$\sigma_r \neq 0$. Reference \cite{straton}, for example, gives a
rigorous analysis on the wave propagation at Fabry-Perot geometry
within the counter-propagating wave approach. In this frame,
additional steps are necessary since the time-averaged Poynting
vector modulus is not derived from the amplitude modulus directly
but from the Poynting vector definition. In turn, because the
S-Formalism uses the radiation intensity as a field variable, its
values at each point within the medium can be directly known and
compared with the squared-field amplitude modulus once the
evolution equations are solved. The solutions $S(z)$ and
$\mathcal{E}^2(z)$ for the Ohmic conductor are derived from
S-formalism evolution equations giving (For resolution details see Appendix
\ref{solapendice}) \small
\begin{subequations}
\begin{widetext}
\begin{eqnarray}
 S(\tilde{z}_r) &=&2\: \frac{\alpha_{_+}^2 \cosh [ \alpha
_{_-}\tilde{d}( \tilde{z}_r-1)] -\alpha _{_+}( \xi+1 ) \sinh [
\alpha _{_-}\tilde{d}( \tilde{z}_r-1) ] +\alpha_{_-}^2 \cos [
\alpha _{_+}\tilde{d}( \tilde{z}_r-1)]-\alpha_{_-} ( \xi -1 ) \sin
[ \alpha_{_+}\tilde{d} ( \tilde{z}_r-1 ) ] }{[
\alpha_{_+}^{2}+(\xi+1 )^2 ] \cosh ( \alpha _{_-}\tilde{d})
+2\alpha _{_+}( \xi+1 ) \sinh ( \alpha _{_-}\tilde{d}) +[
\alpha_{_-}^2 -(\xi -1)^2] \cos ( \alpha_{_+}\tilde{d})+2
\alpha_{_-} ( \xi -1 )\sin (  \alpha_{_+}\tilde{d})},
\label{ec=sc}
\end{eqnarray}
\normalsize and \small
\begin{eqnarray}
\mathcal{E}^2(\tilde{z}_r) &=&4\: \frac{( 1+\xi ) \cosh [ \alpha
_{_-}\tilde{d}(\tilde{z}_r-1) ]-\alpha _{_+}\sinh [ \alpha
_{_-}\tilde{d}( \tilde{z}_r-1) ] +(\xi-1 ) \cos [ \alpha
_{_+}\tilde{d}( \tilde{z}_r-1) ] +\alpha _{_-}\sin [ \alpha
_{_+}\tilde{d}( \tilde{z}_r-1) ]} {[ \alpha_{_+} ^{2}+( \xi +1)^2
] \cosh ( \alpha _{_-}\tilde{d})+2\alpha _{_+}(\xi+1 ) \sinh (
\alpha _{_-}\tilde{d}) +[ \alpha_{_-} ^{2}-( \xi-1 )^2 ] \cos (
\alpha _{_+}\tilde{d}) +2\alpha _{_-}( \xi-1 ) \sin ( \alpha
_{_+}\tilde{d})} ,\label{ec=ec}
\end{eqnarray}
\label{conductor}
\end{widetext}
\end{subequations}

\normalsize \noindent\\[-30pt] \noindent where $\xi = |n_c|$
and $\alpha _{\pm } =2_{Im}^{Re}\{n_{c}\}$ being $n_c$ the complex
refraction index given by $n_c=\sqrt{\epsilon_r+i\sigma_r }$. It
verifies that Eq. (\ref{conductor}) reduces to Eq.
(\ref{eq=soldielectrico}) for $\sigma_r\rightarrow 0$.

\begin{figure}[t]
\includegraphics{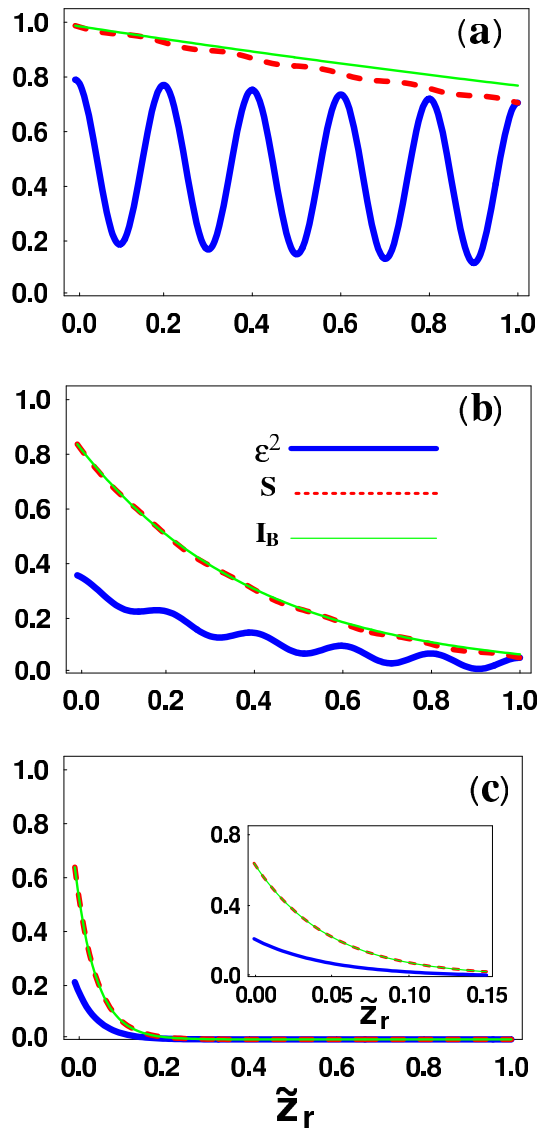}
\caption{(Color online) Spatial evolution of the dimensionless
intensity $S$, squared-field amplitude modulus $\mathcal{E}^2$,
and Beer's Law intensity $I_B$ for a linear absorber with
different relative conductivity values: (a) $\sigma_r=0.1$, (b) 1,
(c) 10, with $\sqrt{\epsilon_r}=2.5$, and $\tilde{d}=2\pi$.
Clearly, the non-equivalence between $S$ and $\mathcal{E}^2$ is
directly related to the effect of the second interface.}
\label{fig=conductor}
\end{figure}

The solutions (\ref{conductor}) are depicted in Fig.
\ref{fig=conductor} for different relative conductivity values
showing that the squared-field amplitude modulus could present a
markedly oscillatory behaviour contrary to the intensity. Figure
\ref{fig=conductor}(a) shows the spatial evolution of both
magnitudes for relatively small values of $\sigma_r$. Both
decrease when the relative spatial coordinate increases due to the
medium absorption. The field attenuation is slight for small
conductivity values which implies an intense contribution to the
total field of the wave returning from the second interface
producing a strong wave superposition. On the other hand, Fig.
\ref{fig=conductor}(b) was depicted for intermediary $\sigma_r$
value. The field attenuation is sufficiently strong and the field
amplitude modulus acquires a quasi-negligible value at interface
$\tilde{z}_r=1$. However, this still produces a wave superposition
so that $\mathcal{E}^2$ shows a slightly oscillatory decreasing
behaviour. Thereby, the intensity and the squared-field amplitude
modulus are yet nonequivalents. Finally, Fig.
\ref{fig=conductor}(c) was depicted for a relatively high value of
$\sigma_r$. In this case, the field amplitude modulus is quite
attenuated before to reach the interface $\tilde{z}_r=1$. Then,
the medium behaves as unbounded and both, the intensity and the
squared-field amplitude, become completely equivalents verifying
\cite{agullo}
\begin{eqnarray}
S(\tilde{z}_r)= \frac{\alpha_+}{2}
\:\mathcal{E}^{2}(\tilde{z}_r).\label{equivalentes}
\end{eqnarray}

In summary, Fig. \ref{fig=conductor} shows that the strength of
wave superposition dynamic is directly related with the effect of
the back interface which turns out in the nonequivalence of $S$
and $\mathcal{E}^2$.

\section{Beer's law comparison}
\begin{figure}[t]
\includegraphics[width=7.5cm,height=5.5cm]{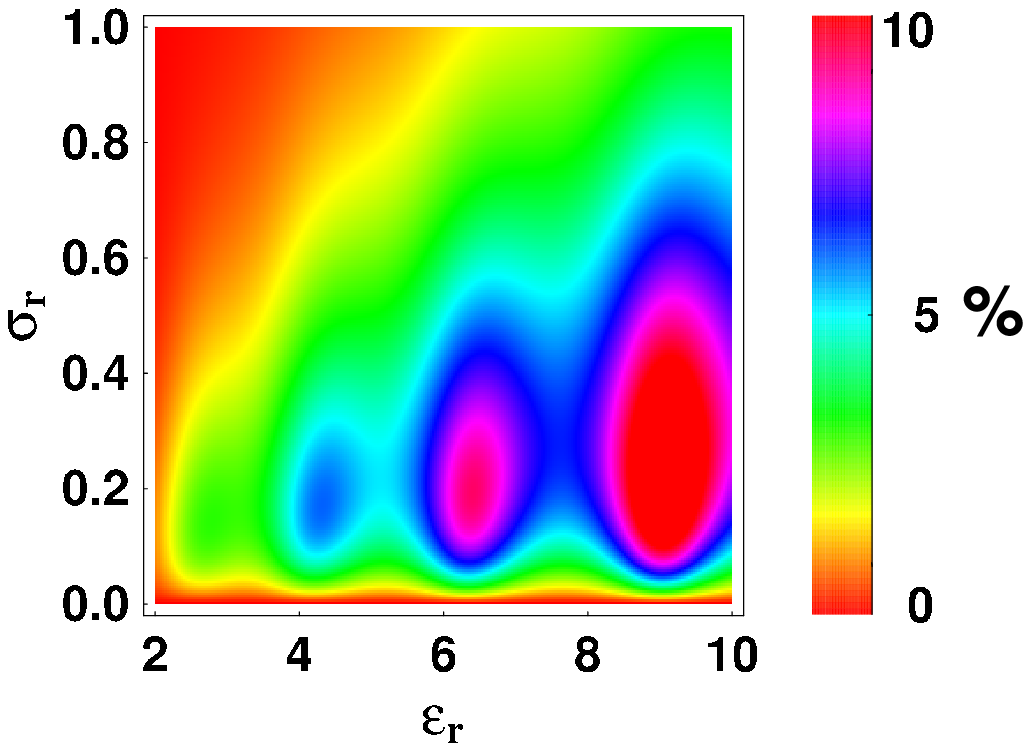}
\caption{(Color online) Percentaged difference $I_B(1)-S(1)$
against $\epsilon_r$ and $\sigma_r$ for $\tilde{d}=2\pi$.}
\label{fig=deferencia}
\end{figure}

The Beer's Law is commonly used to calculate the absorption
coefficient by measuring the transmittance when the optical
absorption within the medium is accounted. This law gives the
intensity of the wave at interface $\tilde{z}_r=1$ when the second
interface effects are neglected \cite{agullo}. Therefore, this Law
is an approximate result since it consider only the first
interface (only one boundary condition) to calculate the energy
flux. The dimensionless intensity given by Beer's Law $I_B$
attenuates exponentially as \cite{agullo}
\begin{eqnarray}
I_B(\tilde{z}_r)=S(0) e^{-\alpha_- \tilde{z}_r},\label{beer}
\end{eqnarray}
where $\alpha_-$ characterize the absorptive medium properties
(note that was used $S(0)$ instead of $I_0$ supposing that the
intensity at $z_r=0$ is know and it lead to diminish the
differences with the exact result). Equation (\ref{beer}) was also
depicted in Fig. \ref{fig=conductor} with the aim to compare the
approximate Beer's Law intensity and the true intensity within the
medium. It happens that $I_B(\tilde{z}_r)$ can departure from
$S(\tilde{z}_r)$ for a wide range of usual parameters. Thereby,
the exact transmittance $S(1)$ could differ from the approximated
transmittance value $I_B(1)$ used often in experiments. It is
clear that a difference appears when the second interface mediates
in the wave dynamic, as Fig. \ref{fig=conductor}(a) shows. When
the effect of that interface is either quasi-negligible [Fig.
\ref{fig=conductor}(b)] or the medium can be considered as
unbounded [Fig. \ref{fig=conductor}(c)], both $I_B(\tilde{z}_r)$
and $S(\tilde{z}_r)$ do not present differences. Thereby, the
validity of this approximation must be carefully tested for each
particular problem since the difference $S(1)-I_B(1)$ depends on
material parameters: $\epsilon_r$, $\sigma_r$ and $\tilde{d}$. In
Fig. \ref{fig=deferencia}, the percentaged difference
$I_B(1)-S(1)$ is shown. As it can see, there exist cases where the
difference of both transmittances is approximately 10$\%$. This
major departure appears when the back interface effect is
relevant\linebreak i. e., when the wave superposition dynamic
plays an important role which happens only for certain particular
values of $(\epsilon_r,\sigma_r)$. Notice that, the maximum
difference appears for high permittivity values and low, but
nonzero, absorption.

\begin{figure*}[t]
\includegraphics[width=18cm,height=4.7cm]{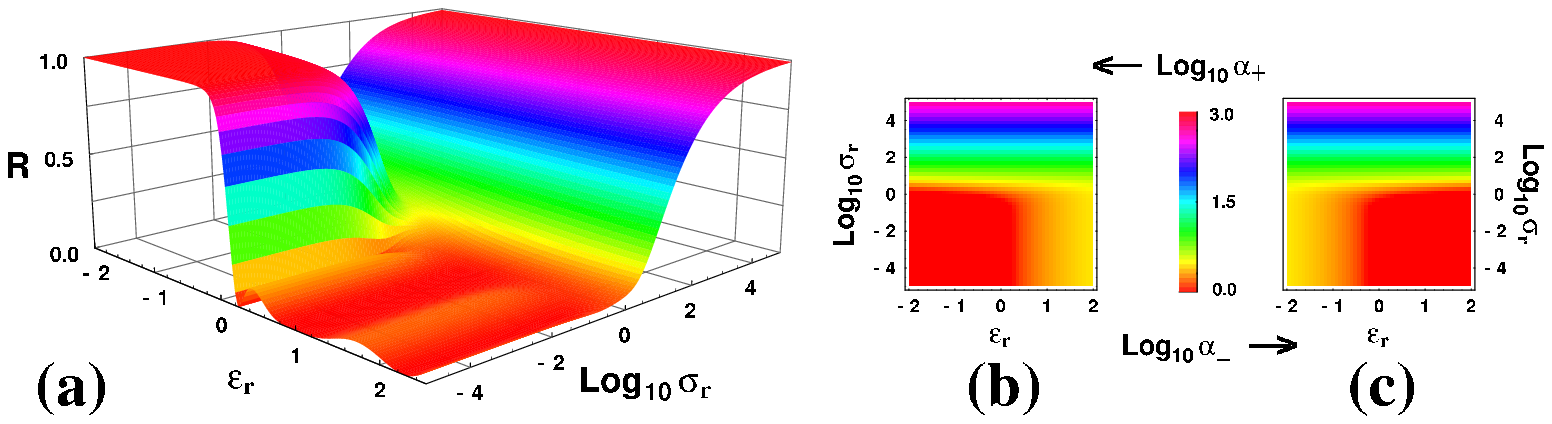}
\caption{(Color online) Reflectance (a), Log$_{10}\alpha_+$ (b),
and Log$_{10}\alpha_-$ (c) against $\epsilon_r$ and
Log$_{10}\sigma_r$. For reflectance $\tilde{d}=2\pi$.}
\label{fig=negativemedia}
\end{figure*}

\section{Reflectance of positive and negative permittivity media}
Negative permittivity media have attracted the attention of
scientist some years ago because was indicated a way to build them
its \cite{pendry} and because was open the possibility of have
others media like the Left-handed Media \cite{science12} leading
to a intense research in this area \cite{2004-06,papel,siteetal}.
However, always the reflectance of negative permittivity media was
analyzed in the context of the effective permittivity of a given
microscopic configuration \cite{pendry,2004-06}. Here we done a
completely macroscopic analysis of the optical properties of
positive and negative permittivity media without enter in the
details of the microscopic configuration.

In Fig. \ref{fig=negativemedia}.(a) the reflectance for both,
positive and negative values of the relative permittivity is shown
for several orders of magnitude of $\sigma_r$. By observing this
figure, we found three regions well differentiated: (1)
$\epsilon_r>0$ and $\sigma_r<1$; (2) $\epsilon_r<0$ and
$\sigma_r<1$; (3)$\sigma_r>1$. Region (1) is characterized by an
oscillating low reflectance whose envelope grows as far as
$\epsilon_r$ increases. This is a well-know behavior
\cite{straton}: the reflectance of a low loss Fabry-P\'erot. On
the other hand, region (2) has a well particular dependence on
$\epsilon_r$ and $\sigma_r$: the reflectance is close to unity.
Finally, region (3) has a uniform reflectance as function of
$\epsilon_r$ that grows monotonously as $\sigma_r$ increases, such
as in the limit $\sigma_r\gg 1$, $R\approx 1$ no matter the
$\epsilon_r$-value.

For a deeper understanding of the reflectance behavior we analyze
the $\alpha_{\pm}$ dependence on $\epsilon_r$ and $\sigma_r$.
Figures \ref{fig=negativemedia}.(b) and
\ref{fig=negativemedia}.(c) depicts $\alpha_+$ and $\alpha_-$,
respectively, as function of $\epsilon_r$ and $\sigma_r$. Also, in
Table I the limiting values for $\sigma_r \ll(\gg)1$ of
$\alpha_{\pm}^2$ and $R$ are shown, to helping in the analysis.
Note that, $\alpha_{\pm}$ represent the \textit{effective}
permittivity and conductivity in all the regions. Therefore, an
analysis of these magnitudes becomes important for a better
physical understanding on the wave dynamic propagation. In fact,
these magnitudes determine the characteristics of the three region
mentioned above, since their values changes dramatically in those.
In region (1), $\alpha_+ \ne 0$ and $\alpha_- = 0$; in region (2)
$\alpha_+ = 0$ and $\alpha_- \ne 0$; and in region (3) $\alpha_+
\approx \alpha_-\ne 0$. Then, the low and oscillatory reflection
correspond to the region where $\alpha_+$ predominates, i.e. the
medium effectively behaves as a dielectric where the oscillations
are produced by the wave superposition as result of the existence
of the second interface summarizing in an Airy-type behavior, as
Table I shows. On the other hand, it is well-known that the high
reflectance is commonly associated with high values of the
conductivity, i.e for good conductors. However, high reflectance
could occur even for low $\sigma_r$ when the permittivity is
negative. In this case, the response is dominated by $\alpha_-$.
The reflectance is given by a hyperbolic-Airy-type function as
Table I show. Because $F'sinh (\delta/2)\gg1$ for almost all the
$\epsilon_r$-values in this region, then $R\approx 1$ and the
medium posses mirror-like properties. For region (3),
$\alpha_{\pm}$ strongly depend on $\sigma_r$-values. The medium
response is completely dominated by $\sigma_r$ with a reflectance
that asymptotically reaches the unity with a moderately slow rate.
Indeed, the asymptotic high reflection region is dictated by a
unique parameter $\alpha \approx \alpha_+ \approx \alpha_-$.
Summarizing, for region (1) is obtained a periodic function of
$\epsilon_r$ for the reflectance, whereas for region (2) it
contains sine hyperbolic functions that monotonously grow as
$\epsilon_r$ decreases achieving rapidly the unity. Note that,
although each reflectance was obtained by calculating the
appropriate limit of Eq. (\ref{conductor}), the reflectance in
region (2) could be obtained from that of region (1) by changing
$\epsilon_r \rightarrow - \epsilon_r$ what allows us to say that
the propagation parameter $\alpha_+$ \textquotedblleft becomes"
imaginary when the permittivity takes negative values doing that
the medium behaves as a high absorbing one.

To end this section we stand out others interesting results: for
$\epsilon_r=1$, $R=0$ because the medium is \textquotedblleft
matched" with the vacuum and all the light is transmitted; For
$\epsilon_r=0$ and $\sigma_r=0$, $R=(1+4/\tilde{d})^{-1}$, showing
that also in this \textquotedblleft quasi-nihility"
\cite{lakhtakia} some quantity of electromagnetic energy can
propagate. And finally, note that for $\epsilon_r<0$ and
$\sigma_r\approx 1$ there exist a region where $R\approx 0.5$
showing that not all the negative permittivity media has a high
reflectance, but those with $\sigma_r \ll 1$. Similar results were
observer for low loss negative permeability media \cite{papel}.

\begin{table}
\begin{ruledtabular}
\caption{Limiting values for $\alpha_{\pm}^2$ and $R$ for the
three regions mentioned in the text.}
\begin{tabular}{c|ccc}
&$\epsilon_r>0$; $\sigma_r\ll 1$&$\epsilon_r<0$; $\sigma_r\ll 1$&$\sigma_r\gg 1$\\
\hline
$\:\:\:\:\:\alpha_+^2\:\:\:\:\:$&$4\epsilon_r$&0&$2\sigma_r$\\[5pt]
$\alpha_-^2$&0&$4|\epsilon_r|$&$2\sigma_r$\\[5pt]
$R$ \footnote{$F'=\frac{(|\epsilon_r|+1)^2}{4|\epsilon_r|}$}
&$\frac{F\sin^2(\delta/2)}{1+F\sin^2(\delta/2)}$&
$\frac{F'\sinh^2(\delta/2)}{1+F'\sinh^2(\delta/2)}$&
$1-2\sqrt{2}/\sigma_r$
\end{tabular}
\end{ruledtabular}
\end{table}

\section{Conclusions}
The S-Formalism was used to study the wave propagation in bounded
linear media. It utilizes the set $(S,\mathcal{E})$ as field
variables instead of the conventional set $(\phi,\mathcal{E})$.
Then, the electromagnetic energy flux can be directly known inside
the medium. The approach stresses that, in general, the
time-averaged Poynting vector modulus and the squared-field
amplitude modulus are non-equivalents. The analysis clearly shows
that the non-equivalence of $S$ and $\mathcal{E}^2$ takes place in
bounded media being consequence of the back interface. The role of
the latter is responsible for the superposition dynamic between
the forward and backward waves which causes an oscillatory
behaviour of the squared-field amplitude modulus contrary to the
intensity. When the second interface can be neglected, the medium
can be treated as unbounded with a single wave propagating such
that $S$ and $\mathcal{E}^2$ are equivalents. The analysis shows
that the usual Beer's Law approximated intensity, employed to
calculate the transmittance in usual experiments, could departure
significatively from exact transmittance and this difference is
also produced by the effect of the second interface. Thereby, the
validity of this approximation should be rigourously tested for
each particular problem. Moreover, positive and negative
permittivity media were analyzed and found different behaviors
accordingly to the permittivity and conductivity values. Was
observed that, for low absorption, negative and positive
permittivity media have a well differentiated reflectivity (being
unity for the former and a low oscillating one for the later)
whereas on the contrary case the reflectance does not depend on
the permittivity values. For intermediate (moderate) absorption,
the reflectance can take values around 0.5, showing that not all
negative permittivity media have high reflectance.

Both methods, the conventional and the S-formalism, should be used
complementarily what could help for a complete physical
understanding on wave propagation in bounded linear media. The
results for finite bandwidth waves can be easily obtained from the
monochromatic ones by integrating all the contributions for the
transmittance \cite{thenext}.

\begin{acknowledgements}
The authors thank Profs. N. Bolognini, H. R. Sandoval and G. M.
Bilmes for helpful suggestions. A.L. thanks to CLAF-CNPq
fellowship.
\end{acknowledgements}

\appendix

\section{True power of a TEM$_{00}$ paraxial Gaussian
beam}\label{appendix}

The free-space propagation of a TEM$_{00}$ Gaussian beam is a good
example on two fundamental aspects: first, because the
misconception about the equivalence of the squared-field amplitude
modulus and the time-averaged Poynting vector exists in the
literature and second, because by applying this misconception
could lead to erroneous physical results.

The theoretical framework on Gaussian beams within the transverse
field-paraxial approximation is well-known \cite{gb,agullo}. The
structure of a TEM$_{00}$ field amplitude, $E(x,y,z)$, propagating
in free-space along $z$-axis with wave vector modulus $k$ (towards
$+z$) is \cite{gb,agullo}
\begin{eqnarray}
E =E_0 \frac{w_{0}}{w\left( z\right) }\exp %
\left[ -\frac{x^{2}+y^{2}}{w^{2}\left( z\right) }\right]\exp
\left[i kz\right]
\:\:\:\:\:\:\:\:\:\:\:\:\:\:\:\:\:\:\:\:\:\:\:\:\:
\nonumber\\[4pt]\:\:\:\:\:\:\:\:\:\:\:
\times\exp %
\left[-i\tan ^{-1}\left( \frac{z}{z_{0}}\right) \right] \exp
\left[ i k\frac{x^{2}+y^{2}}{2R\left( z\right) }\right].
\label{solpara}
\end{eqnarray}
here by simplicity we use dimensional coordinates. The Gaussian
TEM$_{00}$ beam wavefront is perfectly flat at $z=0$ acquiring,
thus, curvature and begin spreading in accordance with
\begin{eqnarray}
R\left( z\right)  =\left[z+ \frac{z_{0}^{2}}{z} \right],\qquad
w^{2}\left( z\right)  =w_{0}^{2}
\left[ 1+\left( \frac{z}{z_{0}}\right) ^{2}%
\right] ,
\end{eqnarray}
where $z_{0} =kw_{0}^{2}/2$, and $z$ is the distance propagated
from the plane $z=0$ where the wavefront is flat being $w_0$ the
minimum spot size often called the waist radius. The parameters
$w(z)$ and $R(z)$ are the spot size and the wavefront radius of
curvature respectively after the wave has propagated a distance
$z$ \cite{gb,agullo}. In Ref. \onlinecite{gb}, it has been
emphasized that the energy per unit time crossing an infinite
transverse plane ($z=cte$) should be a constant to meet the energy
conservation for a losses medium. This calculus has been done
considering equivalents the time-averaged Poynting vector and the
squared-field amplitude modulus since
\begin{subequations}
\begin{eqnarray}
P_{m}=\frac{\varepsilon_{0}\omega}{2 k}\int_{-\infty }^{\infty
}\int_{-\infty }^{\infty }EE^{\ast }\:dxdy
\end{eqnarray}
was used giving the total power \cite{gb}:
\begin{eqnarray}
P_{m}=\frac{\pi\varepsilon_{0}\omega}{4 k}w_{0}^{2}E_{0}^{2},
\end{eqnarray}
\end{subequations}
which effectively is a constant. However, $P_m$ does not represent
the \emph{truly} total power as it will be shown in the following.
Necessarily, the power should be calculated from the time-averaged
Poynting vector as
\begin{subequations}
\begin{eqnarray}
P_{Gb}=\int_{-\infty }^{\infty }\int_{-\infty }^{\infty
}\left\langle S_{z}\right\rangle \:dxdy, \label{energia}
\end{eqnarray}
where $\left\langle S_{z}\right\rangle$ is the $z$-component of
the time-averaged Poynting vector given by Eq. (\ref{sye}).
Calculating (\ref{energia}) it has
\begin{eqnarray}
P_{Gb}=\frac{\pi\varepsilon_{0}\omega}{4
k}w_{0}^{2}E_{0}^{2}\left(1-\frac{1}{k^2 w_0^2}\right),
\label{pot}
\end{eqnarray}
\end{subequations}
so that the power $P_{Gb}$ is different of the power $P_m$
accounted in the literature.
\begin{figure}[t]
\includegraphics[width=8.5cm,height=5.5cm]{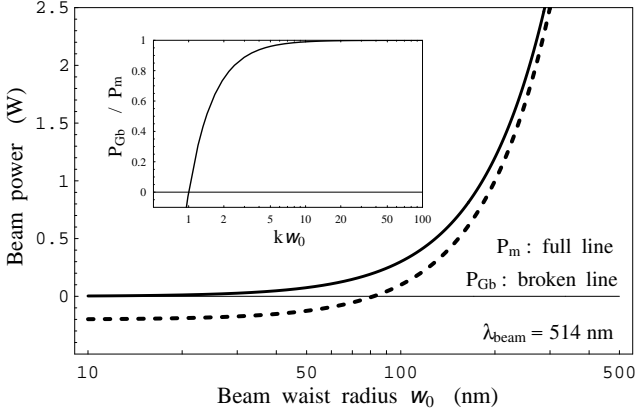}
\caption{Both, $P_m$ and $P_{Gb}$ Gaussian beam powers against
waist radius for $E_0=1.2\times 10^8$ V/m. The inset depicts the
ratio among both powers as function of dimensionless parameter
$kw_0$} \label{gaussian}
\end{figure}

Figure \ref{gaussian} depicts $P_{Gb}$ and $P_m$ against the
minimum spot size simulating an Argon laser beam. In most of the
experimental cases, the beam wavelength is much lesser than that
the beam waist radius satisfying $kw_0\gg1$. In these cases
$P_{Gb}\approx P_m$, as the inset of Fig. \ref{gaussian} shows. It
is clear that both power curves disjoin only from $kw_0 \lesssim
10$. This condition could be satisfied by ultra-focused beams
\cite{dorn}.  In particular, note that, $kw_0\leq1$ leads to
$P_{Gb}\leq 0$ (switch in the beam energy flux direction) which,
of course, is a non-physical result. This unreal value of the
total power $P_{Gb}$ would be consequence of the fact that the
transverse field-paraxial Gaussian beam approach are not longer
valid. Thereby, the power curve $P_{Gb}$ could be useful as
measure parameter to evaluate when this theory can be used: when
the $P_{Gb}$-value is far of the values equal and lesser than
zero, i. e. when $P_{Gb}\approx P_m$. For the example placed in
Fig. \ref{gaussian}, this happens from $w_0\approx 300$ nm. On the
contrary, it is found that always $P_m>0$, which could lead to a
misconception that this value represents the real beam power. The
power curve $P_m$ does not indicate when the theory fails. In
summary, the departure of $P_{Gb}$ from $P_m$ could indicate the
validity degree of the traverse field-paraxial Gaussian beam
approach.

This simples example shows that the misconception about the
equivalence of that magnitudes exists. In the literature, often,
$|E|^2$ is used to calculate the intensity which could lead to
erroneous physical results.

\section{Resolution of Eqs. (\ref{eq=formalismoS})}\label{solapendice}

Here is show the detailed resolution of Eqs.
(\ref{eq=formalismoS}). Thus, the dielectric case is easily
obtained making $\sigma_r=0$.

Equations (\ref{eq=formalismoS}) can be written as:
\begin{subequations}
\begin{eqnarray}
2\frac{d^2u}{d \tilde{z}^2}u-\left(\frac{d u}{d
\tilde{z}}\right)^2+
4\epsilon_r u^{2}=4S^{2},\label{eq=apformS1}\\
\frac{d S}{d \tilde{z} } =-\sigma_r u,
\end{eqnarray}
\label{eq=apformS12}
\end{subequations}
\noindent whit $u=\mathcal{E} ^{2}$ and the boundary conditions
reads
\begin{subequations}
\begin{eqnarray}
\left[\frac{d u}{d \tilde{z}}\right]_{\tilde{z}=\tilde{d}} &=&0,\\
\left[S- u\right]_{\tilde{z}=\tilde{d}} &=&0,\\
\left[
(u+S)^2+\frac{1}{4}\frac{du}{d\tilde{z}}-4u\right]_{\tilde{z}=\tilde{d}}
&=&0.
\end{eqnarray}
\end{subequations}

\emph{By the homogeneity of the Eqs. (\ref{eq=apformS12})}, we
write down the following \textit{ansatz}:
\begin{subequations}
\begin{eqnarray}
u=A\exp\left( a \tilde{z}\right),\\
v=B\exp\left( a\tilde{z}\right).
\end{eqnarray}
\end{subequations}
Replacing in Eq. (\ref{eq=apformS12}), it has
\begin{eqnarray}
a^{2}=\pm \alpha_{\mp}.
\end{eqnarray}
Then, the general solution can be written (\emph{by the
homogeneity of Eqs. ( \ref{eq=apformS12})}) as a linear
combination of the four possible values of $a$:
\begin{subequations}
\begin{eqnarray}
u=A_{1}exp\left( \alpha_- \tilde{z}\right) +A_{2}exp\left(
-\alpha_- \tilde{z}\right)+\:\:\:\:\qquad\qquad\qquad\nonumber\\[4pt]
+A_{3}exp\left( i \alpha_+ \tilde{z}\right)+ A_{4}exp\left( -i
\alpha_+ \tilde{z}\right)\qquad\qquad\label{eq=solu}
\end{eqnarray}
\begin{eqnarray}
v=\frac{\alpha_+\alpha_-}{2} \left[- \frac{A_{1}}{\alpha_-
}exp\left( \alpha_- \tilde{z}\right)
+\frac{A_{2}}{\alpha_-}exp\left( -\alpha_- \tilde{z}\right)+
\right.\qquad\nonumber\\[4pt]
\left. +i\frac{A_{3}}{\alpha_+ }exp\left( i k_0\alpha_+
\tilde{z}\right) -i\frac{A_{4}}{\alpha_+ }exp\left( -i \alpha_+
\tilde{z}\right) \right]\qquad.
\end{eqnarray}
\end{subequations}

The boundary conditions fix three of the four constants appearing
in the solution, the fourth is fixed by auto-consistency.
Replacing Eq. (\ref{eq=solu}) in Eq. (\ref{eq=apformS1}), it
results
\begin{eqnarray}
A_{1}A_{2}=A_{3}A_{4}.
\end{eqnarray}
Applying the boundary conditions at $\tilde{z}=\tilde{d}$ and
defining $B=A_{1}/A_{3}$, $ C=A_{2}/A_{3}$, results
$BC=A_{4}/A_{3}$, has
\begin{subequations}
\begin{eqnarray}
B&=&\left[ \frac{2i\xi-\left( \alpha_- +i\alpha_+ \right)
}{2i\xi+\left( \alpha_- +i\alpha_+ \right)}\right] \exp \left[
\left( -\alpha_- +i\alpha_+ \right) \tilde{d}%
\right],\qquad\\[4pt]
C&=&\left[ \frac{2i\xi-\left( \alpha_- -i\alpha_+ \right)
}{2i\xi+\left( \alpha_- -i\alpha_+ \right)}\right] \exp \left[
\left( \alpha_- +i\alpha_+ \right) \tilde{d}%
\right],
\end{eqnarray}
\end{subequations}
and with the boundary condition at $\tilde{z}=0$
\begin{eqnarray}
A_{3}=4\big[B\left( 1+\xi-\alpha_+ \right) +C\left(1+\xi+\alpha_+
\right)+\nonumber\\[5pt]+
BC\left(1-\xi-i\alpha_- \right)  +\left( 1-\xi+i\alpha_-
\right) \big]^{-1}.
\end{eqnarray}

By replacing this values in Eq. (\ref{eq=solu}), it gives\\[-33pt]
\onecolumngrid
\noindent\\[1pt]
--------------------------------------------------------------------------$\!^{|}$

\begin{subequations}
\begin{eqnarray}
u=4\:\frac{ ( 1+\xi -\alpha _{_+}) \exp\left[ \alpha _{_-}(
\tilde{z}_r-1)\right] +( 1+\xi +\alpha _{_+}) \exp\left[ -\alpha
_{_-}( \tilde{z}_r-1)\right] +} {( 1+\xi -\alpha _{_+})^{2}\exp[
-\alpha _{_-}\tilde{d}] +( 1+\xi +\alpha _{_+})^{2}\exp[ \alpha
_{_-}\tilde{d}] -}\qquad\qquad\qquad\qquad\qquad\qquad\qquad\qquad\qquad\nonumber\\[12pt]
\frac{+( \xi -1+i\alpha _{_-}) \exp\left[ -i \alpha _{_+}(
\tilde{z}_r-1)\right]+( \xi -1-i\alpha _{_-}) \exp\left[ i\alpha
_{_+}( \tilde{z}_r-1)\right]}{-( \xi -1+i\alpha _{_-}) ^{2}\exp[ i
\alpha _{_+}\tilde{d}]-( \xi -1-i\alpha _{_-}) ^{2}\exp[ -i
\alpha_{_+}\tilde{d}]},\qquad
\end{eqnarray}

\noindent and

\begin{eqnarray}
 v =2\:\frac{ ( 1+\xi +\alpha _{_+}) \alpha
_{_+}\exp\left[ -\alpha _{_-}( \tilde{z}_r-1)\right] -( 1+\xi
-\alpha _{_+}) \alpha _{_+}\exp\left[ \alpha _{_-}(
\tilde{z}_r-1)\right]+ }{( 1+\xi -\alpha _{_+}) ^{2}\exp[ -\alpha
_{_-}\tilde{d}] +( 1+\xi +\alpha _{_+}) ^{2}\exp[ \alpha
_{_-}\tilde{d}] -
}\qquad\qquad\qquad\qquad\qquad\qquad\qquad\qquad
\nonumber\\[12pt]
\frac{ +i( \xi -1-i\alpha _{_-}) \alpha _{_-}\exp\left[ i \alpha
_{_+}( \tilde{z}_r-1)\right] -i( \xi -1+i\alpha _{_-}) \alpha
_{_-}\exp\left[ -i \alpha _{_+}(\tilde{z}_r-1)\right] }{ -( \xi
-1+i\alpha _{_-}) ^{2}\exp[ i \alpha _{_+}\tilde{d}] -( \xi
-1-i\alpha _{_-}) ^{2}\exp[ -i \alpha _{_+}\tilde{d}]}\qquad
\end{eqnarray}
\end{subequations}

\noindent\\[5pt]
\noindent From this equations, Eqs. (\ref{conductor}) can be
obtained rearranging the exponential terms. \twocolumngrid

\end{document}